# Giant field-tunable nonlinear Hall effect by Lorentz skew scattering in a graphene moiré superlattice

Pan He[1,2†*], Min Zhang[1†], Yue-Xin Huang[3,4†], Jingru Li[1], Ruibo Wang[1], Shiwen Zhao[1], Chaoyu Pan[1], Yuxiao Gao[1], Takashi Taniguchi[5], Kenji Watanabe[6], Junxiong Hu[7], Yinyan Zhu[1,2*], Cong Xiao[8*], X. C. Xie[2,8,9], Shengyuan A. Yang[10] and Jian Shen[1,2,11,12,13,14*]

[1]*State Key Laboratory of Surface Physics and Institute for Nanoelectronic devices and Quantum computing, Fudan University, Shanghai 200433, China*

[2]*Hefei National Laboratory, Hefei 230088, China*

[3]*School of Sciences, Great Bay University, Dongguan 523000, China*

[4]*Great Bay Institute for Advanced Study, Dongguan 523000, China*

[5]*Research Center for Materials Nanoarchitectonics, National Institute for Materials Science, 1-1 Namiki, Tsukuba 305-0044, Japan*

[6]*Research Center for Electronic and Optical Materials, National Institute for Materials Science, 1-1 Namiki, Tsukuba 305-0044, Japan*

[7]*School of Physics, University of Electronic Science and Technology of China, Chengdu 611731, China*

[8]*Interdisciplinary Center for Theoretical Physics and Information Sciences (ICTPIS), Fudan University, Shanghai 200433, China*

[9]*International Center for Quantum Materials, School of Physics, Peking University, Beijing 100871, China*

[10]*Research Laboratory for Quantum Materials, Department of Applied Physics, The Hong Kong Polytechnic University, Hong Kong, China*

[11]*Department of Physics, Fudan University, Shanghai, China*

[12]*Shanghai Research Center for Quantum Sciences, Shanghai, China*

[13]*Zhangjiang Fudan International Innovation Center, Fudan University, Shanghai 201210, China*

[14]*Collaborative Innovation Center of Advanced Microstructures, Nanjing 210093, China*

[†]These authors contributed equally to this work.
[*]Correspondence to: hepan@fudan.edu.cn; zhuyinyan@fudan.edu.cn; congxiao@fudan.edu.cn; shenj5494@fudan.edu.cn

**Abstract:**
**The nonlinear Hall effect (NHE) can enable rectification and energy harvesting, and its control by external fields, including gate, strain and magnetic field, has been pursued intensively. However, existing tuning pathways rely predominantly on fully quantum mechanical effects and are typically inefficient, resulting in weak NHE signals that limit further progress. In this**

**work, we report the discovery of a distinct type of NHE in a graphene–hBN moiré superlattice, which arises from a classical-quantum cooperative effect called Lorentz skew scattering (LSK), induced by a perpendicular magnetic field. This field-driven NHE exhibits a linear dependence on magnetic field and a pronounced unidirectional angular dependence. Remarkably, its magnitude reaches up to 32% of the linear Hall signal. We show that this giant, field-tunable NHE originating from LSK follows a unique quartic scaling law and produces a record-high nonlinear Hall conductivity (36000 μmV$^{-1}$Ω$^{-1}$) near van Hove singularities of moiré minibands, which is over an order of magnitude larger than all previously reported NHEs. Our findings establish an efficient, magnetic-field-driven route to giant Hall rectification in high-mobility materials, offering a broadly applicable paradigm for modulating the NHE beyond electrostatic gating.**

**Introduction:**

The study of Hall effects is a cornerstone of condensed matter physics, driving both fundamental discoveries and technological innovations[1]. Recently, a new member of the Hall family, the nonlinear Hall effect (NHE), has emerged as a higher-order response to an applied current[2]. The NHE has been observed in both non-magnetic[3-8] and magnetic[9,10] materials, with mechanisms attributed to Berry curvature dipole (BCD)[11], nonlinear side jump and skew scatterings[2,12-14], Berry connection polarizability (BCP)[15], and nonlinear Drude transport[16]. To enhance and modulate the NHE, which are critical for future rectification and energy-harvesting applications, various external tuning pathways have been explored, including magnetic fields[16-20], gate voltages[3,7-10], and strain[21,22]. These approaches reveal rich quantum-control behaviors of electronic band and wave functions. However, because they rely on subtle quantum-mechanical perturbations, their tuning effect is usually weak and unable to greatly enhance the NHE. For example, in previous experiments, magnetic fields modified the BCP[20,23] or nonlinear Drude mechanisms[16] through Zeeman coupling corrections to wave functions and bands, producing relatively small NHE signals. This limitation motivates the search for new paradigms of field-driven NHE, in which the external field acts as a central ingredient of the NHE mechanism rather than a minor perturbation to already known zero-field NHE mechanisms. Ideally, such a mechanism would provide both strong tunability and substantially enhanced nonlinear response.

A recent theoretical work introduced a novel mechanism, Lorentz skew scattering (LSK), which arises from the cooperative action of the Lorentz force and skew scattering (Fig. 1a)[24]. In highly conductive materials, the classical Lorentz force effect dominates over quantum perturbation effects of magnetic field. Indeed, this LSK mechanism predicts a distinctive NHE wherein the nonlinear Hall conductivity scales *quartically* with the longitudinal conductivity ($\sigma_{xx}^4$), in stark contrast to the $\sigma_{xx}^2$ or $\sigma_{xx}^0$ scaling expected from Zeeman-modified Drude[16] or intrinsic BCP[20,23] mechanisms. A higher scaling power implies a stronger NHE in high-conductivity systems, suggesting great potential for efficient frequency doubling or rectification applications[5,6]. Notably, the strongest zero-field NHE reported to date, arising from nonlinear skew scattering[7,8], exhibits only cubic scaling, highlighting the unique potential of LSK to produce an even larger effect. Despite its theoretical promise, experimental evidence for LSK has been lacking.

Skew-scattering-induced NHE has been already investigated both theoretically[12,13] and experimentally[5,7,8]. In particular, large zero-field NHE induced by skew scattering has been observed in various broken inversion symmetry graphene[7,8]. Such symmetry breaking can be realized, for example, by aligning graphene with hexagonal boron nitride (hBN), which opens a gap at the Dirac point and induces valley-contrasting Berry curvature[25,26]. While Berry curvature drives skew scattering under an in-plane electric field **E**[12,13,24], a perpendicular magnetic field **H** induces an additional Lorentz force in two-dimensional (2D) systems. This raises the intriguing possibility that the combined action of **E** and **H** could induce a previously unobserved, magnetic-field-tunable NHE, as proposed theoretically[24].

Here, we report the experimental observation of a magnetic-field-tunable NHE originating from LSK in a high-mobility graphene-hBN moiré superlattice. The observed NHE exhibits a characteristic dependence on the magnetic and electric fields ($\propto E^2 H \cos\theta$, where θ is the magnetic field angle relative to the sample normal), distinct from all known mechanisms. The effect is significantly enhanced near van Hove singularities of moiré minibands, where Berry curvature hotspots coincide with peaks in density of states. The nonlinear Hall conductivity scales quartically with $\sigma_{xx}$, confirming the LSK origin through our theoretical analysis. This mechanism enables efficient magnetic-field control of the NHE's magnitude and polarity, achieving a record-high nonlinear Hall conductivity of 36000 μmV$^{-1}$Ω$^{-1}$. Our findings establish a new paradigm for achieving magnetic-field-driven, tunable, and giant NHEs in high-mobility quantum materials.

**Results:**

To realize high-quality monolayer graphene with broken inversion symmetry, we fabricated hBN/graphene/hBN heterostructures (Fig. 1b), in which the top hBN was precisely aligned with the graphene lattice and the bottom hBN was intentionally misaligned. This stacking configuration not only breaks the graphene's inversion symmetry but also generates a long-period moiré superlattice (~ 14 nm) for the perfectly-aligned interface due to the small (~1.8%) lattice mismatch between graphene and hBN (Fig. 1b)[27-29]. The hBN-encapsulated graphene exhibits a high carrier mobility (~100,000 cm$^2$/V·s), making it an ideal platform for probing second-order nonlinear transport effects arising from skew scattering[7,8,12] and LSK[24]. We fabricated Hall bar devices using electron beam lithography and reactive ion etching, followed by deposition of Cr/Au (3 nm/90 nm) contacts. A representative device is shown in Fig. 1b. The moiré potential significantly reconstructs the graphene band structure, creating moiré minibands, including low-energy secondary Dirac points (DPs)[28,29] and van Hove singularities[7,30,31]. These features appear in the longitudinal resistance $R_{xx}$ and the Hall resistance $R_{xy}$ as a function of the gate voltage $V_g$ (Figs. 1c,1d)[7,30]. Here, the subscripts $x$ and $y$ indicate the current direction and the direction transverse to current, respectively. The distinct peaks in $R_{xx}$ at $V_g \approx 0$ V and $V_g \approx 34$ V corresponds to the primary and secondary DPs, respectively, which are accompanied by sign reversals in $R_{xy}$ under a magnetic field. While the existence of Berry curvature in this system[25,26,32] is crucial for facilitating skew scattering, the role of external magnetic fields in this process has been largely overlooked. This gap in understanding motivates our investigation of field-induced NHE in this high-mobility moiré platform.

**Observation of a field-tunable NHE**

We measured the second harmonic transverse voltage $V_y^{2\omega}$ under an AC current of frequency ω to probe the NHE. This work focuses on $V_y^{2\omega}$ induced by an external magnetic field, so the zero-field contribution has been subtracted from the data to better visualize the field-driven effect. We observed a pronounced, field-induced $V_y^{2\omega}$ in Fig. 1e. The magnitude of $V_y^{2\omega}$ varies dramatically with $V_g$ and changes sign with reversing the magnetic field (**H**) direction. Moreover, $V_y^{2\omega}$ exhibits a quadratic current dependence and switches sign upon simultaneously reversing the directions of current and voltage probes (Supplementary Fig. 1), consistent with the nature of second-order NHE. To obtain the field-odd component of the NHE, we calculated $\Delta V_y^{2\omega} = [V_y^{2\omega}(+H) - V_y^{2\omega}(-H)]/2$. As shown in Fig. 1f, $\Delta V_y^{2\omega}$ exhibits pronounced peaks near both the primary and secondary DPs. The gate dependence of $\Delta V_y^{2\omega}$ differs markedly from that of the ordinary Hall effect (Fig. 1d), indicating distinct underlying mechanisms. The graphene-hBN moiré superlattice has three-fold rotational symmetry, thus both BCP and BCD contributions to the NHE are symmetry-forbidden[10,11,33]. The NHE observed here can be governed by the interplay between Lorentz force and skew scattering[24], while the linear Hall effect arises from the Lorentz force. According to the LSK mechanism, the NHE additionally depends on the Berry curvature distribution and density of states at the Fermi surface[24] (Methods), both of which vary significantly with Fermi level in the graphene moiré superlattice[25,26]. This accounts for the complex gate dependence observed in $\Delta V_y^{2\omega}$.

To avoid complications arising from Landau level formation, all our measurements were conducted under relatively low magnetic fields ($\leq 0.5$ T). Within this range, we find that $V_y^{2\omega}$ increases linearly with *H* (Supplementary Fig. 2). To investigate the directional dependence of the field-induced NHE, we measured $V_y^{2\omega}$ while rotating the magnetic field out of the sample plane. As shown in Fig. 1g, $V_y^{2\omega}$ follows a cosine-like angular dependence with a 360° period. The amplitudes $\Delta V_y^{2\omega}$, extracted from the cosine fits, are plotted as a function $V_g$ in Supplementary Fig. 3, showing consistency with the trends observed in Figs. 1e and 1f. This cosθ angular dependence is consistent with the vector nature of the Lorentz force in a 2D system: the effect maximizes for perpendicular fields and vanishes when the field is in-plane.

The unidirectional angular dependence of the NHE was further observed at different magnetic fields ranging from +0.2 T to -0.2 T, as shown in Fig. 2a. The extracted amplitude $\Delta V_y^{2\omega}$ increases linearly with *H* (Fig. 2b). These findings reinforce the essential role of the Lorentz force. The measurements under different currents (Fig. 2c) show that $\Delta V_y^{2\omega}$ scales quadratically with current amplitude (Fig. 2d), further confirming the second-order nature of the NHE. Together, the above findings demonstrate that the observed NHE follows a characteristic $E^2H\cos\theta$ dependence, distinct from previously reported NHE, which typically scale as $E^2H^0$ [3-10,16] or $EH$ [15]. Notably, this field-induced NHE does not require the presence of spin-orbit coupling, which is however a central ingredient in many earlier NHE studies[3-6]. Our findings establish an effective control knob to engineer NHE in graphene moiré superlattices.

**Temperature dependence and quartic scaling**

To investigate the temperature dependence of the field-induced NHE, we measured the $V_y^{2\omega}(\theta)$ curves at various temperatures (Fig. 3a). As shown in Fig. 3b, the $\Delta V_y^{2\omega}$ decreases dramatically with rising temperature, a trend consistently observed across different $V_g$ values, such as $V_g$ = -5 V and 50.5 V. In contrast, the linear Hall signal $\Delta V_y^{1\omega}$ exhibits weak temperature dependence (Fig. 3c). Moreover, the NHE maintains the same sign at the two different $V_g$ (Fig. 3b), while the linear Hall effect reverses sign (Fig. 3c). These contrasting behaviors reflect their distinct physical origins. Notably, the magnitude of the NHE reaches up to 32% of the linear Hall effect at low temperatures (Fig. 3d).

In addition, the nonlinear Hall conductivity $\sigma_{yxx}$, calculated using $\sigma_{yxx} = \frac{2\sigma_{xx}V_y^{2\omega}L^2}{(V_x^{1\omega})^2 W}$, exhibits a significant temperature dependence (Fig. 4a). To confirm the origin of the NHE, we analyzed the relationship between $\sigma_{yxx}$ and $\sigma_{xx}$ by plotting $\sigma_{yxx}$ as a function of $\sigma_{xx}$ (Fig. 4b). A dominant quartic scaling $\sigma_{yxx} \propto \sigma_{xx}^4$ is observed from the fitting, which differs from previous reports of NHE and represents the highest power-law dependence observed to date. This scaling strongly supports LSK as the dominant mechanism behind the NHE. Remarkably, we observed a giant $\sigma_{yxx}$ value up to 390 μmV$^{-1}$Ω$^{-1}$ in the graphene-hBN moiré superlattice Device 1 (Fig. 4c). The gate dependence of $\sigma_{yxx}$ ($V_g$) shows multiple peaks located near van Hove singularities, where $\sigma_{xx}(V_g)$ exhibits maxima (Fig. 4d)[7,30,31,34]. These regions correspond to enhanced density of states and Berry curvature "hot spots"[34], which can significantly amplify the LSK contribution[24] (Methods).

The above findings were consistently reproduced across multiple graphene-hBN moiré superlattice devices with varying moiré wavelengths and thus Fermi energies of van Hove singularity[30,31,34] (Supplementary Figs. 4-7). In all devices, we observed a robust field-induced NHE, characterized by an $E^2H\cos\theta$ dependence and a dominant $\sigma_{yxx} \propto \sigma_{xx}^4$ scaling. While the overall gate dependence of $\sigma_{yxx}$ remained qualitatively similar among these devices, their magnitude varied a lot mainly due to their differences in carrier mobility (or $\sigma_{xx}$). The quartic scaling ($\sigma_{yxx} \propto \sigma_{xx}^4$) implies that high $\sigma_{xx}$ devices naturally host giant nonlinear transport effects. Notably, the Device 2 (Device 3), which exhibits approximately twice (sevenfold) the $\sigma_{xx}$ of Device 1, shows a record-high $\sigma_{yxx}$ of 1600 (36000) μmV$^{-1}$Ω$^{-1}$ with an overwhelmingly dominant quartic scaling behavior (Supplementary Figs. 4f and 6f)[7,8,35]. In contrast, field-induced NHE is not expected in graphene with preserved inversion symmetry, where the Berry curvature vanishes. To verify this, we fabricated non-aligned hBN/graphene/hBN heterostructures, where the inversion symmetry remains intact. As anticipated, these control devices exhibited a negligible nonlinear Hall response (Supplementary Fig. 8). This stark contrast further underscores the critical role of inversion symmetry breaking and Berry curvature in enabling the LSK mechanism.

**Theoretical understandings**

In the following, we further confirm that the LSK is the physical origin of our observation through theoretical transport analysis. As shown in Ref. 24, the LSK current of $E^2B$ order is given by (Methods)

$$\boldsymbol{j}^{LSK} = e\tau^4 \sum_l \boldsymbol{v}_l [\hat{D}_E\{\hat{D}_L, \hat{I}_{sk}\} + \hat{D}_L\{\hat{I}_{sk}, \hat{D}_E\} + \hat{I}_{sk}\{\hat{D}_E, \hat{D}_L\}]\hat{D}_E f^0 , \qquad (1)$$

where $\boldsymbol{v}_l$ is the group velocity of a Bloch state with a composite band-momentum index $l = (n, \boldsymbol{k})$, $f^0$ is the equilibrium Fermi-Dirac distribution, $\hat{D}_E = -e\boldsymbol{E} \cdot \partial_{\boldsymbol{k}}$ and $\hat{D}_L = -e\boldsymbol{v}_l \times \boldsymbol{H} \cdot \partial_{\boldsymbol{k}}$ are differential operators corresponding to field driving terms in the Boltzmann transport equation, whereas $\hat{I}_{sk}$ is the skew scattering integral operator, whose action on a distribution function $f_l$ reads $\hat{I}_{sk}f_l = -\sum_{l'} \omega_{l'l}^{3a}(f_l + f_{l'})$, with $\omega_{l'l}^{3a}$ as the skew scattering rate (details in Methods). The notation $\{\hat{D}_E, \hat{D}_L\}$ denotes the anticommutator of two operators. At finite temperatures, the relaxation time $\tau$ is contributed by both impurities and phonons, whereas the skew scattering rate $\omega_{l'l}^{3a}$ is only contributed by impurity scattering[7,36]. As such, when the temperature is varied, $\boldsymbol{j}^{LSK} \sim \tau^4 \sim \sigma_{xx}^4$ manifests a quartic scaling, which is consistent with our experimental observation. Moreover, we present model calculation results that provide further compelling evidence for the LSK mechanism. By using $\sigma_{xx} = 0.01$ S from experiment and assuming a chemical potential near the band edge $\mu = 20$ meV, we estimate the density of Coulomb impurities is about $n_i \approx 2 \times 10^{10}$ $m^{-2}$ (dielectric constant $\varepsilon \approx 3$). The second-order conductivity from LSK, under an applied magnetic field $H = 0.5 \, T$, is then calculated to be $\sigma_{yxx} \approx 58$ μmV$^{-1}$Ω$^{-1}$ (see Methods), being in reasonable agreement with the experimental observation near the primary DP.

Next, we exclude other nonlinear transport mechanisms with quartic scaling, suggesting that the LSK is indeed the predominant effect. In a nonmagnetic material, the time-reversal-even ($\mathcal{T}$-even) current response of $E^2H$ order can arise from two categories (I and II) of contributions. In category I, the magnetic field enters through the classical Lorentz force effect (Fig. 1a), which combines with quantum processes of first order $\mathcal{T}$-odd electrical transport[37,38], including Berry curvature anomalous velocity, side jump and skew scattering. The scaling form of this category can be evaluated as follows: the Lorentz force effect gives $EH\tau^2$ scaling as familiar in Drude-Boltzmann transport theory[39], whereas in temperature varying measurements the $\mathcal{T}$-odd linear electrical transports manifest scaling forms of $E\tau^0$, $E\tau^1$, and $E\tau^2$ [36]. Combining them, the current response of $E^2H$ order thus takes the scaling form of $E^2H\tau^2$, $E^2H\tau^3$, and $E^2H\tau^4$. Notably, the $E^2H\tau^4$ scaling comes from LSK, which dominates over other terms in highly conductive materials with large $\tau$. In category II, the magnetic field enters through quantum mechanical effects by Zeeman coupling to electronic magnetic moment[40,41] and by coupling to Berry curvature (such as the k-space density-of-states correction[42]). These perturbation effects modify the $\mathcal{T}$-odd zero-$H$-field NHE mechanisms, giving rise to $\mathcal{T}$-even transport of $E^2H$ order. In temperature varying measurements, the $\mathcal{T}$-odd electrical nonlinear transport has scaling forms of $E^2\tau^i$, with $i = 0, 1, 2, 3, 4$,[43] and the corresponding $\mathcal{T}$-even transport of $E^2H$ order takes the scaling as $E^2H\tau^i$, where the quartic-scaling terms come from the compositions of two skew scattering processes (SKSK)[43]. We thus need compare the LSK with SKSK. First, we remind that a more detailed scaling analysis showed that[24] $j^{LSK} \sim \tau^4/\tau^{sk}$ and $j^{SKSK} \sim \tau^4/(\tau^{sk})^2$ since the $\mathcal{T}$-odd skew-scattering linear transport $\sim \tau^2/\tau^{sk}$, where $\tau^{sk}$ roughly measures the time scale of skew scattering, which is much larger than $\tau$.[5,7,12] Because $1/\tau^{sk} \sim \omega^{3a} \sim n_i$, with $n_i$ being the impurity concentration, $j^{LSK}$ should be much larger than $j^{SKSK}$ in high-mobility materials with small $n_i$. Second, we perform model calculations of Zeeman corrected SKSK contribution (Methods), and the results indeed show that it is much smaller than LSK. In addition, we note that this comparison can also be anticipated in highly conductive systems, because the effect of category I is from classical-quantum mixture (e.g., LSK from mixture of Lorentz force and skew scattering), which naturally dominates over the purely quantum effect in category II (e.g., Zeeman corrected SKSK).

The above theoretical analysis strongly supports the LSK mechanism for the observed quartic scaling of NHE. As a side remark, the dominance of category I contribution in highly conductive materials also suggests that the observed $E^2H\tau^3$ scaling found in Device 1 is likely to be accounted for by Lorentz side jump mechanism. We remind that this $E^2H\tau^3$ scaling cannot arise from a purely classical effect due to the cooperation of Lorentz force induced normal Hall transport ($EH\tau^2$ scaling) and linear Drude transport ($E\tau^1$ scaling) because this contribution is $\mathcal{T}$-odd (Methods) thus cannot appear in nonmagnetic materials.

**Discussion:**

In summary, we report the discovery of a novel type of NHE in graphene-hBN moiré superlattices, induced by a perpendicular magnetic field. This NHE exhibits a unidirectional angular dependence on the magnetic field and reverses sign upon field reversal. The nonlinear Hall voltage scales quadratically with the applied current and linearly with the magnetic field, demonstrating a distinct $E^2H\cos\theta$ dependence. Our combined experimental and theoretical study confirms that this NHE is induced by the interplay between classical Lorentz force and quantum skew scattering, resulting in a nonlinear Hall conductivity that scales quartically with longitudinal conductivity. This novel LSK mechanism yields a record-high nonlinear Hall conductivity of 36000 μmV$^{-1}$Ω$^{-1}$ in graphene moiré superlattice with tunable low-energy van Hove singularities, which is over an order of magnitude larger than all previously reported nonlinear Hall conductivities. Furthermore, the ability to control and manipulate the NHE using an external magnetic field opens exciting opportunities for novel device applications, including tunable Hall rectifiers and low-power nonlinear signal processors.

As an outlook, the magnetic-field-tunable NHE reported in this work is expected to be broadly generalizable across a variety of 2D materials, including bilayer graphene, few layer graphene and transition metal dichalcogenides. Large NHE responses are anticipated near band crossings, where both the Berry curvature and the density of states are enhanced. Systems with flat electronic bands, including twisted bilayer graphene[44], twisted multilayer graphene[45], and rhombohedral multilayer graphene[46-48], are particularly promising, as they host strong correlation effects and pronounced Berry curvature "hot spots". In these systems, field-tunable NHE may serve as a powerful experimental probe of Berry curvature near the Fermi surface. The LSK effect is not limited to nonlinear electrical transport but is also expected to play a significant role in other nonlinear transport phenomena, such as nonlinear thermal and thermoelectric Hall transports. In particular, it may lead to strong thermal Hall rectification[49] and significant nonlinear Nernst effect[50] in graphene superlattices, which are promising research directions. Looking beyond 2D systems, this new type of NHE could also extend to three-dimensional inversion-breaking materials, where electrostatic gating is challenging but magnetic field control remains viable. Future investigations along this direction may also uncover new aspects of quantum transport under high magnetic fields, including regimes characterized by Shubnikov-de Haas[51] and Brown-Zak oscillations[52,53].

**Methods**

**Device fabrication**

The graphene-hBN moiré superlattices were fabricated using a conventional dry-transfer technique[54]. First, hBN and graphene flakes were mechanically exfoliated onto separate Si/SiO$_2$ (285 nm) substrates. Suitable hBN flakes (20-40 nm thick) and monolayer graphene were identified by optical microscopy and monolayer graphene were further verified by Raman spectroscopy. Next, the assembly was carried out with a polycarbonate (PC) film supported on a polydimethylsiloxane (PDMS) stamp. The top hBN layer was first picked up by the PC/PDMS stamp, followed by the monolayer graphene, whose crystal axes were carefully aligned (typically by matching straight edges) to the hBN to form the moiré superlattice. The bottom hBN layer was then picked up to complete the heterostructure stack. The entire stack was released onto a highly doped Si substrate at the PC melting temperature of 180 °C. The full width at half maximum of the graphene Raman 2D peak was used to estimate the twist angle between graphene and hBN.

For electrical transport measurements, Hall-bar devices were fabricated from a clean region in the above heterostructures using a two-step electron-beam lithography process combined with reactive ion etching (RIE). A bilayer resist stack of MMA EL6 and PMMA A4 was spin-coated to define the etch mask. The exposed regions were etched using $CHF_3/O_2$ plasma, and residual resists were removed by solvent cleaning. Subsequently, electron-beam lithography was used again to pattern the electrode contacts, followed by thermal evaporation of Cr/Au (3 nm/90 nm) layers. The resulting devices exhibited low contact resistance and clean interface, suitable for high-mobility transport measurements.

**Electrical transport measurements**

Electric transport measurements were performed in a Physical Properties Measurement System (PPMS, Quantum Design) equipped with a rotatable sample holder. We conducted AC measurements, where a sinusoidal current $I^\omega = I\sin\omega t$ with a frequency $\omega/2\pi$=17.777 Hz was applied to Hall-bar devices using a current source meter (Keithley 6221). To simultaneously probe the linear and nonlinear transport responses, the first- and second-harmonic longitudinal and Hall voltages were measured by multiple lock-in amplifiers (Stanford Research SR830). The phases of lock-in amplifiers were set to 0° and 90° for the first and second harmonic voltage measurements, respectively. A sub-femtoamp source meter (Keithley 6430) was employed to apply the back-gate voltage and monitor leakage currents. Magnetic fields of variable magnitude and orientation were applied during the measurements to investigate field-dependent transport behaviors.

**LSK formalism.**

The expression of LSK can be derived from Boltzmann equation. For the homogeneous low-frequency case, the Boltzmann equation reads

$$\dot{\boldsymbol{k}} \cdot \partial_{\boldsymbol{k}} f_l = \hat{I}_{el}\{f_l\}, \qquad (2)$$

where $\hat{I}_{el}\{f_l\}$ denotes the collision integral and $f_l$ is the distribution function. Typically, the collision integral can be divided into three terms: $\hat{I}_{el} = \hat{I}_c + \hat{I}_{sj} + \hat{I}_{sk}$, where $\hat{I}_c f_l = -\sum_{l'} \omega^s_{l'l}(f_l - f_{l'})$ is the conventional collision integral, $\hat{I}_{sj}$ is the side jump collision integral, and $\hat{I}_{sk}f_l = -\sum_{l'} \omega^a_{l'l}(f_l + f_{l'})$ is the skew-scattering collision integral. $\hat{I}_{sj}$ is irrelevant to LSK, and will not be further considered in the following. Here, $\omega^s_{l'l} = (\omega_{l'l} + \omega_{ll'})/2$ is the symmetric part of scattering rate $\omega_{ll'}$ between $l$ and $l'$ states, and $\omega^a_{l'l} = (\omega_{l'l} - \omega_{ll'})/2$ is the anti-symmetric part. The scattering rate is given by the golden rule $\omega_{ll'} = \frac{2\pi}{\hbar}|T_{ll'}|^2\delta(\epsilon_l - \epsilon_{l'})$, where $T_{ll'}$ is the scattering $T$ matrix determined by the Lippmann-Schwinger equation, and $\epsilon_l$ is the band energy. The magnetic field and electric field enter transport through the equations of motion

$$\dot{\boldsymbol{k}} = -\frac{e}{\hbar}\boldsymbol{E} - \frac{e}{\hbar}\dot{\boldsymbol{r}} \times \boldsymbol{B}, \qquad (3)$$

$$\dot{\boldsymbol{r}} = \partial_{\hbar \boldsymbol{k}}\varepsilon_l - \dot{\boldsymbol{k}} \times \boldsymbol{\Omega}_l. \qquad (4)$$

The noncanonical structure of the equations of motion also leads to a corrected k-space density of states $\mathfrak{D} = 1 + \mathbf{\Omega}_l \cdot \mathbf{B}$. Moreover, the band energy has a correction from Zeeman coupling to spin and orbital magnetic moment. As we pointed out in the main text, transport contributions of category I dominate in highly conductive systems. To capture this category of contribution, it is sufficient to retain the Boltzmann equation in the form of[24] $(\widehat{D}_E + \widehat{D}_L)f_l = (\hat{I}_c + \hat{I}_{sk})f_l$, where $\widehat{D}_E = -\frac{e}{\hbar}\mathbf{E} \cdot \partial_{\mathbf{k}}$ and $\widehat{D}_L = -\frac{e}{\hbar}(\mathbf{v}_l \times \mathbf{B}) \cdot \partial_{\mathbf{k}}$, with $\mathbf{v}_l = \partial_{\hbar\mathbf{k}}\varepsilon_l$ being the group velocity. The LSK distribution function, order of $\sim E^2 B$, can be obtained by acting two $\widehat{D}_E$'s, one $\widehat{D}_L$ and one $\hat{I}_{sk}$ on the Fermi-Dirac distribution $f^0$. Taking the relaxation time approximation for the conventional part of collision integral $\hat{I}_c f_l \approx -f_l/\tau$, and summing over all possible sequences of the four relevant operators and noting that $\hat{I}_{sk}$ and $\widehat{D}_L$ cannot directly generate nonequilibrium distribution from $f^0$, one reaches the expression of LSK current presented in Eq. (1)[24]. In addition, the purely classical contribution $\mathbf{j} = \tau^3 \sum_l \mathbf{v}_l \{\widehat{\mathcal{D}}_E, \widehat{\mathcal{D}}_L\}\widehat{\mathcal{D}}_E f_0$ from the combination of Lorentz force induced normal Hall transport ($EH\tau^2$ scaling) and linear Drude transport ($E\tau^1$ scaling) vanishes in nonmagnetic materials, because the integrand of this contribution is an odd function of $\mathbf{k}$ under time reversal symmetry.

For spin-independent impurities, the leading order skew scattering rate takes the form of
$$\omega^{3a}_{ll} \approx \frac{4\pi^2}{\hbar} n_i \sum_{l''} \langle V_{ll''} V_{l''l'} V_{l'l} \rangle_c \delta(\varepsilon_{l'} - \varepsilon_l)\delta(\varepsilon_{l''} - \varepsilon_l) \mathrm{Im}\, W(l, l', l''),$$
where $n_i$ is the impurity density, $V_{ll}$ is the scattering potential, $\langle ... \rangle_c$ denotes the disorder average, and $W(l, l', l'') = \langle u_l | u_{l'} \rangle \langle u_{l'} | u_{l''} \rangle \langle u_{l''} | u_l \rangle$ is the Wilson loop connecting the three involved electronic states $l$, $l'$, and $l''$. Therefore, the skew scattering is nonzero only if the Pancharatnam-Berry phase $\arg(W)$ is nonzero. Interestingly, for an infinitesimal Wilson loop in $k$ space, one finds[15] $\mathrm{Im}\, W(l, l', l'') \approx \frac{1}{2}(\mathbf{k}'' - \mathbf{k}) \times (\mathbf{k}' - \mathbf{k}) \cdot \mathbf{\Omega}_l$. These characters indicate that skew scattering and thus the LSK benefits from strong Berry curvature. Moreover, from Eq. (1), one sees that there are three-fold momentum integrals over $\mathbf{k}''$, $\mathbf{k}'$, and $\mathbf{k}$ in evaluating the LSK current. Large density of states around the Fermi surface is thus also favored for enhanced LSK effect.

**Effective model of monolayer graphene with inversion symmetry breaking.**
The effective model of graphene near the Dirac point takes the form of[7,12]
$$H_s = [svk_x - \lambda(k_x^2 - k_y^2)]\sigma_x + (vk_y + 2s\lambda k_x k_y)\sigma_y + \Delta\sigma_z, \tag{5}$$
where $s = \pm 1$ denote $K/K'$ valleys in the Brillouin zone, and $\sigma_i$'s are Pauli matrices in the sublattice space. The hBN substrate breaks inversion symmetry and induces an energy gap $\Delta$. $\lambda$ denotes trigonal warping, which is consistent with the $C_3$ rotation symmetry of graphene/hBN and is manifested by k-quadratic terms around $K/K'$ valleys.

We consider the Coulomb scattering source $V(\mathbf{r}) = \sum_j \frac{e^2 Q}{4\pi\varepsilon_0\varepsilon|\mathbf{r}-\mathbf{r}_j|}$, where $\varepsilon_0$ is the vacuum permittivity, $\varepsilon$ is the dielectric constant, $\mathbf{r}_j$ is the position of the scattering source, and $eQ$ is its

charge. To evaluate the LSK current in graphene, we first calculate the relaxation time with

$$\frac{1}{\tau} = \sum_{l'} \omega_{l'l}^{s}(1 - \cos\langle \boldsymbol{k}, \boldsymbol{k'}\rangle) = \left(\frac{e^2 Q}{4\pi\varepsilon_0\varepsilon}\right)^2 \frac{n_i \pi^2}{\hbar} \frac{\mu^2 + \Delta^2}{\mu(\mu^2 - \Delta^2)},$$

by assuming $\lambda$ is small enough.

We estimate the conductivity when $v/\hbar = 0.94 \times 10^6$ m/s, $\lambda = va/4$, $a = 1.42$ Å, the mass gap $\Delta = 15$ meV, $Q = 1$, and dielectric constant $\varepsilon \approx 3$. By employing the eigenvectors of Eq. (5), the linear longitudinal conductivity can be expressed as $\sigma_{xx} = \frac{e^2 \tau}{\hbar^2} \frac{\mu^2 - \Delta^2}{4\pi\mu}$ when the chemical potential $\mu$ intersects the energy bands. Then, using an observed value $\sigma_{xx} = 0.01$ S from experiment and assuming a chemical potential near the band edge $\mu = 20$ meV, we estimate the impurity density is about $n_i \approx 2 \times 10^{10}\, m^{-2}$, and the second-order conductivity under an applied magnetic field $B = 0.5\, T$ is $\sigma_{yxx} \approx 58\, \mu\text{m} \cdot \text{V}^{-1}\Omega^{-1}$, which reasonably agrees with the experimental observations near the primary DP. We also calculate the Zeeman coupling corrected SKSK contribution[24,43] $\boldsymbol{j}^{SKSK} = -e\tau^4 \sum_l \boldsymbol{v}_l [\widehat{D}_E \hat{I}_{sk} \hat{I}_{sk} + \hat{I}_{sk} \widehat{D}_E \hat{I}_{sk} + \hat{I}_{sk} \hat{I}_{sk} \widehat{D}_E] \widehat{D}_E f^0$ to the NHE, where we choose the g-factor to be 100, and find that it is more than three orders of magnitude smaller than the LSK.

**Acknowledgments**

P.H. was sponsored by National Key Research and Development Program of China (grant no. 2022YFA1403300), National Natural Science Foundation of China (grant no.12174063 and U23A2071), Natural Science Foundation of Shanghai (23ZR1403600) and the start-up funding from Fudan University. This work was supported by Innovation Program for Quantum Science and Technology grant 2024ZD0300103. Y.H was supported by National Natural Science Foundation of China (grant no.12474041) and the Pearl River Talent Recruitment Program (2023QN10X746). C.X. was sponsored by National Natural Science Foundation of China (grant no.12574114) and the start-up funding from Fudan University. S.A.Y. was supported by The HK PolyU Start-up Grant No. (P0057929). X.C.X. was sponsored by Quantum Science and Technology - National Science and Technology Major Project (Grants No. 2021ZD0302400). K.W. and T.T. acknowledge support from the JSPS KAKENHI (Grant Numbers 21H05233 and 23H02052), the CREST (JPMJCR24A5), JST and World Premier International Research Center Initiative (WPI), MEXT, Japan.

**Author contributions**

P.H., Y.Z. and J.S. planed the study. P.H. and M.Z. performed the transport measurements and the data analysis. M.Z., J.L., R.W., S.Z. C.P. and Y.G and J. H. fabricated devices. T.T. and K.W. provided the hBN crystal. Y.H., C.X. X.C.X and S.A.Y. performed theoretical studies. P.H., M.Z., Y.H., Y.Z., C.X., and J.S. wrote the manuscript with the input from all authors. All authors commented the manuscript.

**Competing interests** The authors declare no competing interests.

**Data availability**

The data that support the findings of this study are available within the paper and the Supplementary Information. Other relevant data are available from the corresponding authors upon reasonable request. Source data are provided with this paper.

**Code availability**

The codes that support this study are available from the corresponding author upon reasonable request.

**Figures**

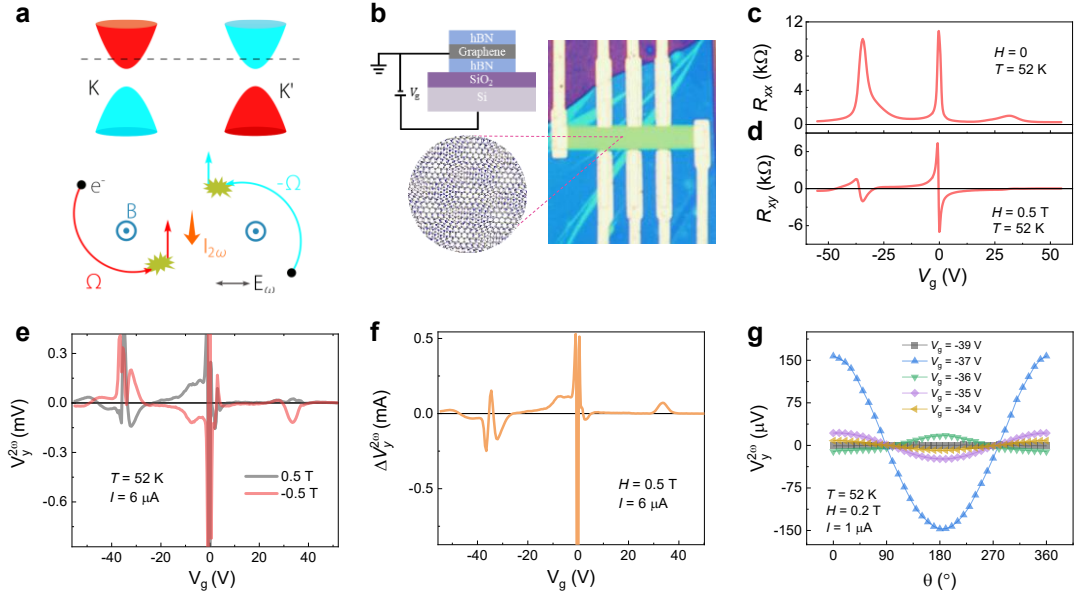

**Fig. 1| Magnetic field-induced nonlinear Hall effect (NHE) in a graphene-hBN moiré superlattice via Lorentz skew scattering. a**, Schematic of Lorentz skew scattering (LSK) in gapped graphene. The upper panel shows the two valleys in the Brillouin zone, represented by gapped Dirac cones with opposite Berry curvatures. The lower panel illustrates the mechanism: A perpendicular magnetic field induces circular motion in opposite directions for Bloch electrons in the two valleys. Due to the opposite Berry curvatures, LSK causes electrons from both valleys to deflect into the same transverse direction, resulting in a non-zero net transverse signal. **b**, Schematic illustrations of hBN-encapsulated monolayer graphene (top left) and the resulting moiré superlattice (bottom left), alongside a microscopic photograph of a typical Hall bar device (right panel). **c, d**, Longitudinal resistance ($R_{xx}$) under zero magnetic field (**c**) and Hall resistance ($R_{xy}$) under $H = 0.5$ T (**d**) as a function of back gate voltage ($V_g$). **e**, Second harmonic transverse voltage $V_y^{2\omega}$ as a function of $V_g$ for two opposite magnetic fields ($H = +0.5$ T and -0.5 T). **f**, Field-odd component of the second harmonic transverse voltage $\Delta V_y^{2\omega}$ as a function of $V_g$. **g**, $V_y^{2\omega}$ as a function of field angle θ for several different $V_g$ at $I = 1$ μA and $H = 0.2$ T. Th solid lines represent cosθ fits to the experimental data. θ is defined as the angle between the sample normal and the magnetic field direction.

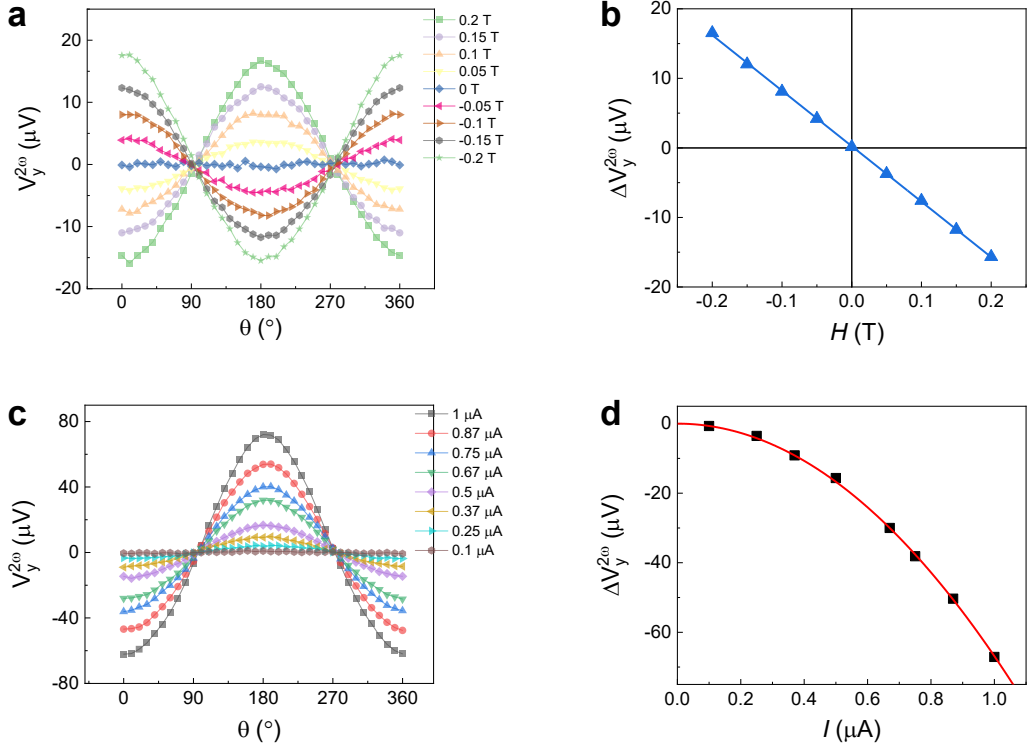

**Fig. 2| Magnetic field and current dependences of $V_y^{2\omega}$. a**, $V_y^{2\omega}$ as a function of θ for several different magnetic field strengths at $I$ = 0.5 μA. A vertical offset has been subtracted from the data to better visualize the field-driven effect. **b**, $\Delta V_y^{2\omega}$ as a function of magnetic field. The solid line represents a linear fit to the data. **c**, $V_y^{2\omega}$ as a function of θ for several different currents at $H$ = 0.2 T. A vertical offset has been subtracted from the data to emphasize the field-driven effect. **d**, $\Delta V_y^{2\omega}$ as a function of current. The red line shows a quadratic fit to the data. All results in this figure were measured at $T$ = 50 K and $V_g$ = -36 V.

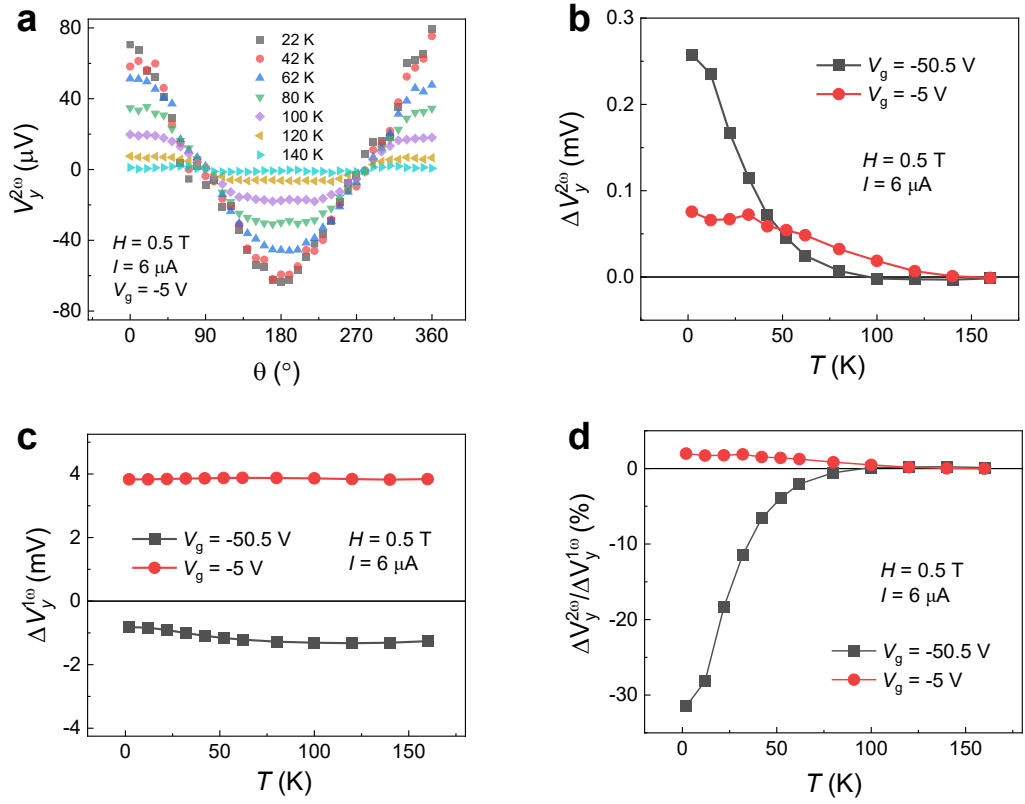

**Fig. 3| Temperature dependence of field-induced NHE. a**, $V_y^{2\omega}$ as a function of θ for several different temperatures. A vertical offset has been subtracted from the data to emphasize the field-driven effect. **b**, $\Delta V_y^{2\omega}$ as a function of $T$ for two representative $V_g$. **c**, $\Delta V_y^{1\omega}$ as a function of $T$ for two representative $V_g$. **d**, The ratio between $\Delta V_y^{2\omega}$ and $\Delta V_y^{1\omega}$ as function of $T$ for the same two representative $V_g$.

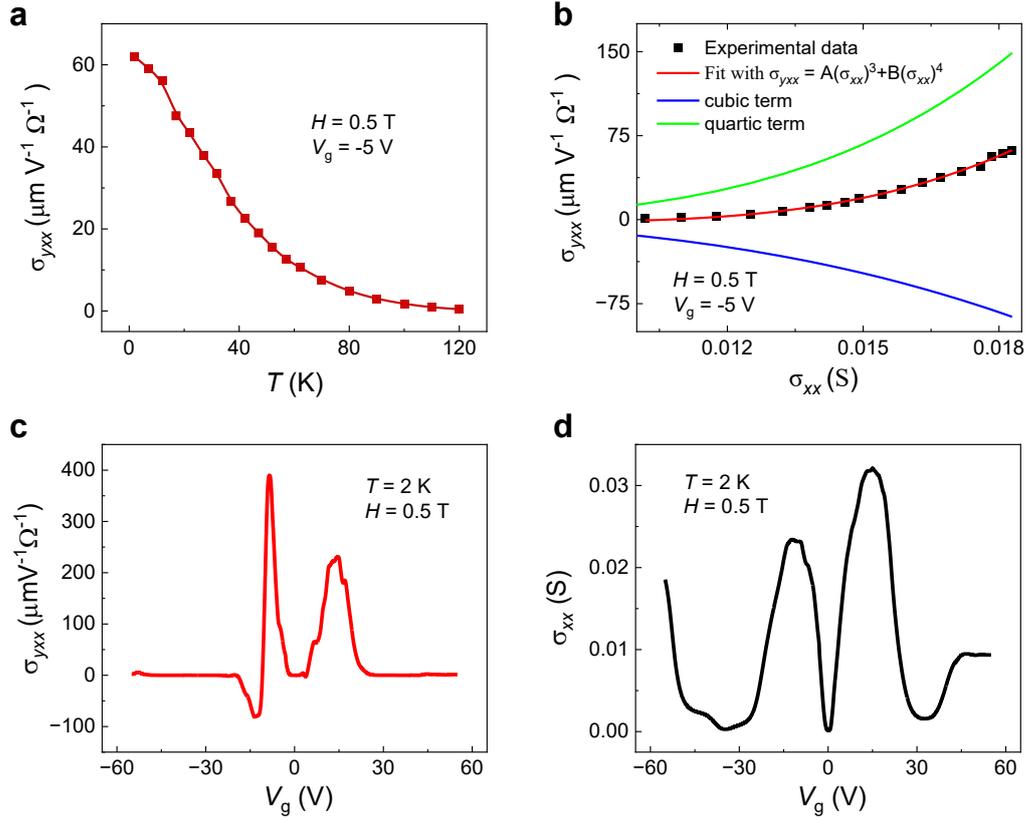

**Fig. 4| The scaling law of NHE and its temperature and gate voltage dependences. a**, The nonlinear Hall conductivity $\sigma_{yxx}$ as a function of temperature. **b**, $\sigma_{yxx}$ as a function of $\sigma_{xx}$. The red line is a fit to the formula $\sigma_{yxx} = A(\sigma_{xx})^3 + B(\sigma_{xx})^4$. $A$ and $B$ are fitting parameters. The contributions from the cubic and quartic terms are also plotted in the figure. **c**, $\sigma_{yxx}$ as a function of $V_g$. **d**, $\sigma_{xx}$ as a function of $V_g$. The results in **a-d** were obtained at $V_g = $ -5 V.

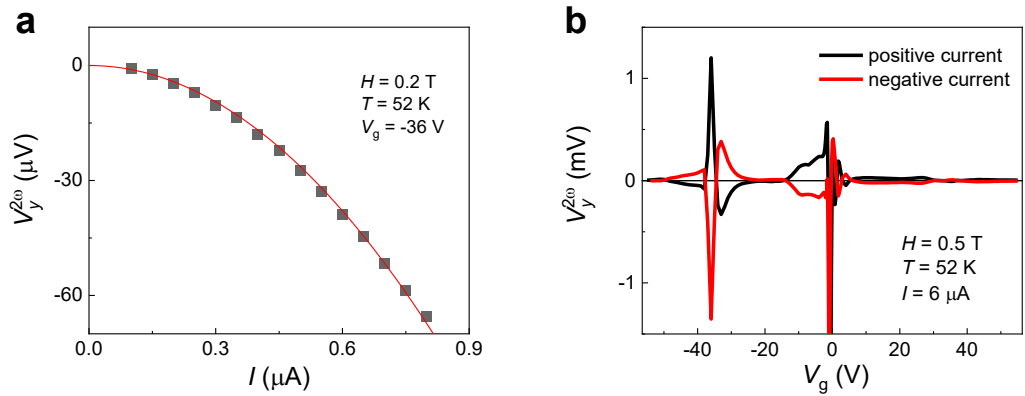

**Supplementary Fig. 1| The quadratic current dependence of $V_y^{2\omega}$. a**, $V_y^{2\omega}$ as a function of current. The red line represents a quadratic fit to the data. **b**, $V_y^{2\omega}$ as a function of $V_g$ under two opposite currents. The voltage probes are switched simultaneously with reversing current. These results were obtained at $H = $ 0.2 T, $T = $ 52 K and $V_g = $ -36 V in moiré superlattice Device 1.

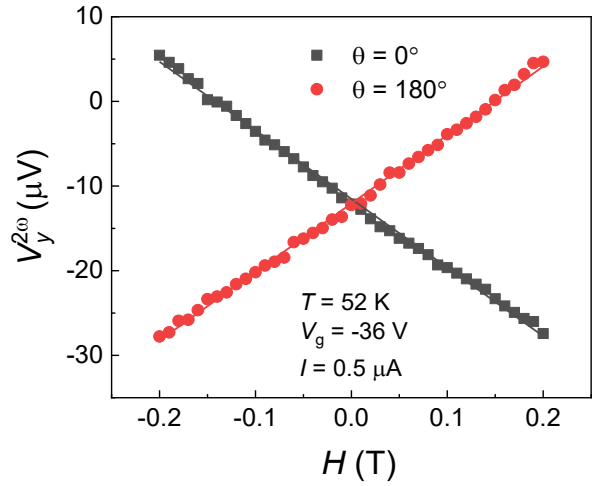

**Supplementary Fig. 2| The linear magnetic field dependence of $V_y^{2\omega}$.** $V_y^{2\omega}$ as a function of magnetic field $H$ for the $\theta = 0°$ and $180°$. The solid lines show linear fittings to the data. These results were obtained at $T = 52$ K, $V_g = -36$ V and $I = 0.5$ μA in moiré superlattice Device 1.

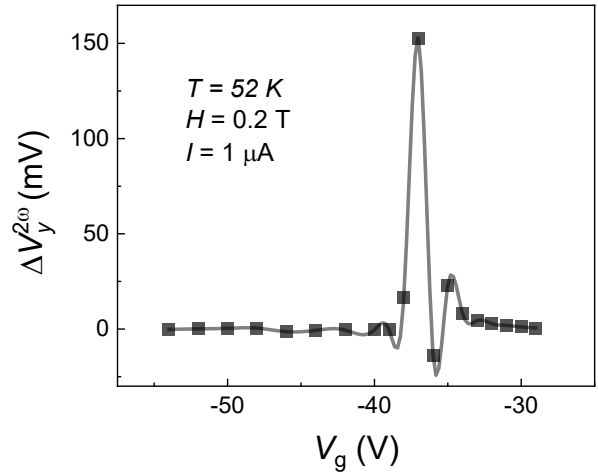

**Supplementary Fig. 3| The gate voltage dependence of $\Delta V_y^{2\omega}$.** The $\Delta V_y^{2\omega}$ extracted from the cosine fits to data in Fig. 1e are plotted as a function $V_g$

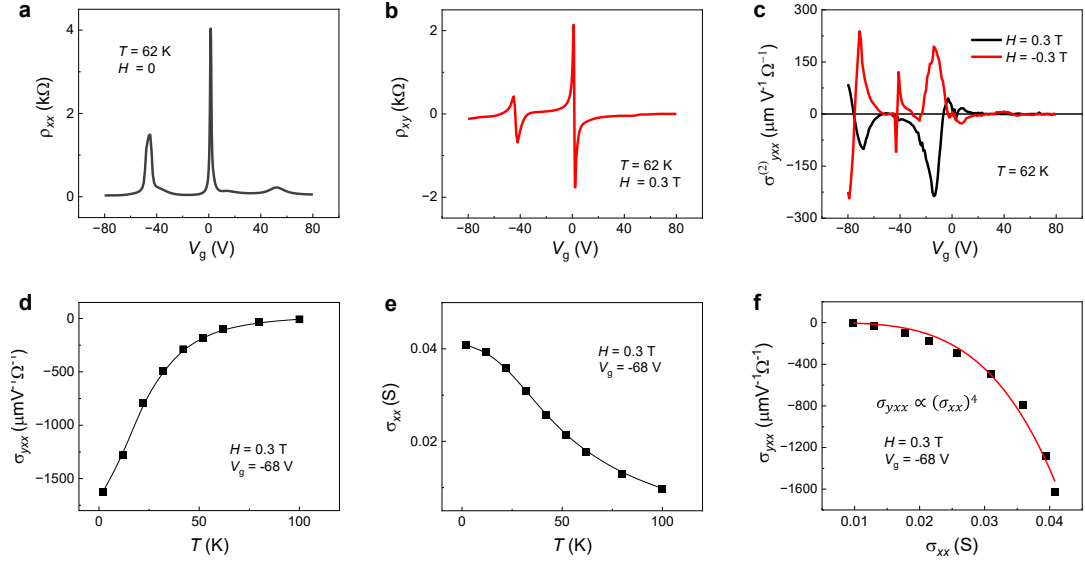

**Supplementary Fig. 4| The field-induced NHE in moiré superlattice Device 2. a**, The longitudinal resistivity ρ$_{xx}$ as a function of $V_g$ under zero magnetic field. **b**, The Hall resistance ρ$_{xy}$ as a function of $V_g$ under $H$ = 0.3 T. **c**, The nonlinear Hall conductivity $\sigma_{yxx}$ as a function of $V_g$. **d**, $\sigma_{xx}$ as a function of temperature. **e**, $\sigma_{yxx}$ as a function of temperature. **f**, $\sigma_{yxx}$ as a function of $\sigma_{xx}$. The red line represents a fit to the formula $\sigma_{yxx} = B(\sigma_{xx})^4$.

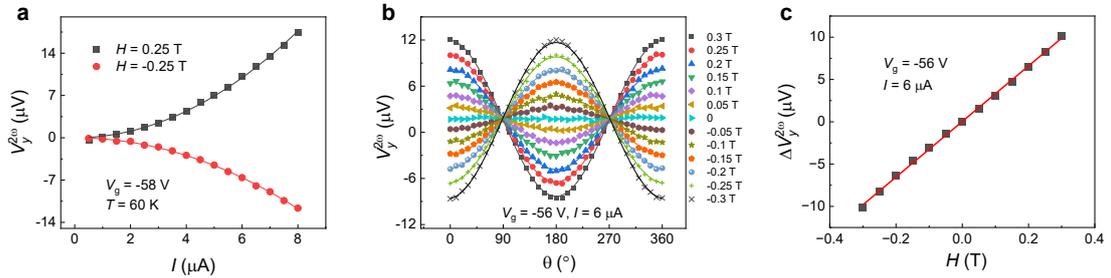

**Supplementary Fig. 5| The current $I$, field angle θ and strength $H$ dependences of NHE in moiré superlattice Device 2. a**, $V_y^{2\omega}$ as a function of $I$ at two opposite magnetic fields. The solid lines represent quadratic fittings to the data. **b**, $V_y^{2\omega}$ as a function of θ for several different magnetic fields. A finite $V_y^{2\omega}$ at zero magnetic field has not been subtracted from the data. **c**, $\Delta V_y^{2\omega}$ as a function of magnetic field. The solid line represents a linear fit to the data.

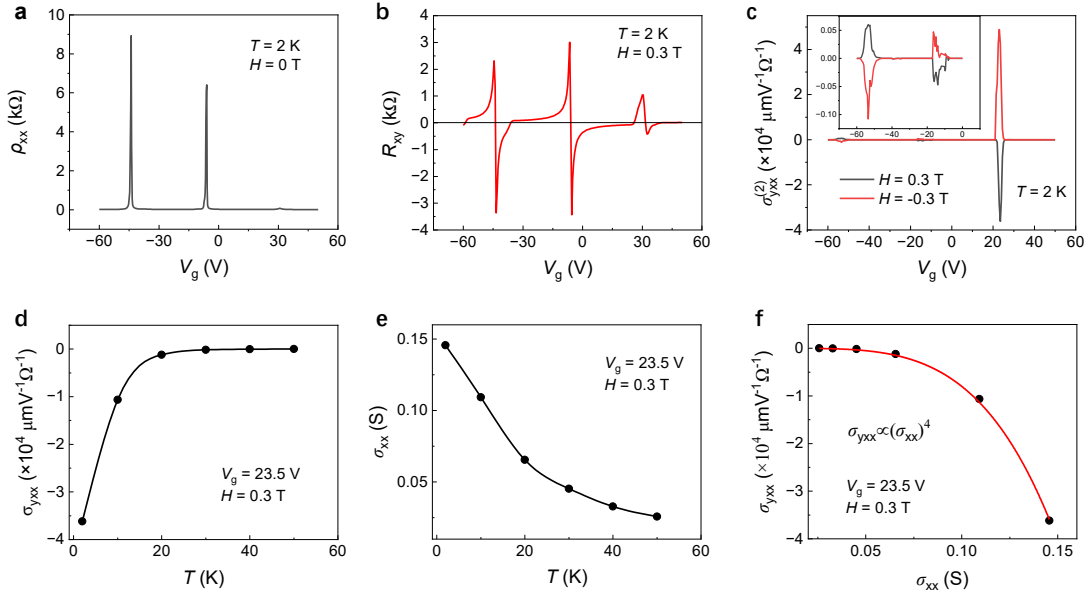

**Supplementary Fig. 6| The field-induced NHE in moiré superlattice Device 3.** **a**, The longitudinal resistivity $\rho_{xx}$ as a function of $V_g$ under zero magnetic field. **b**, The Hall resistivity $\rho_{xy}$ as a function of $V_g$ under $H$ = 0.3 T. **c**, The nonlinear Hall conductivity $\sigma_{yxx}$ as a function of $V_g$. The inset shows the magnified plot covering the $V_g$ from 0 to -60 V. **d**, $\sigma_{yxx}$ as a function of temperature. **e**, $\sigma_{xx}$ as a function of temperature. **f**, $\sigma_{yxx}$ as a function of $\sigma_{xx}$. The red line represents a fit to the formula $\sigma_{yxx} = B(\sigma_{xx})^4$.

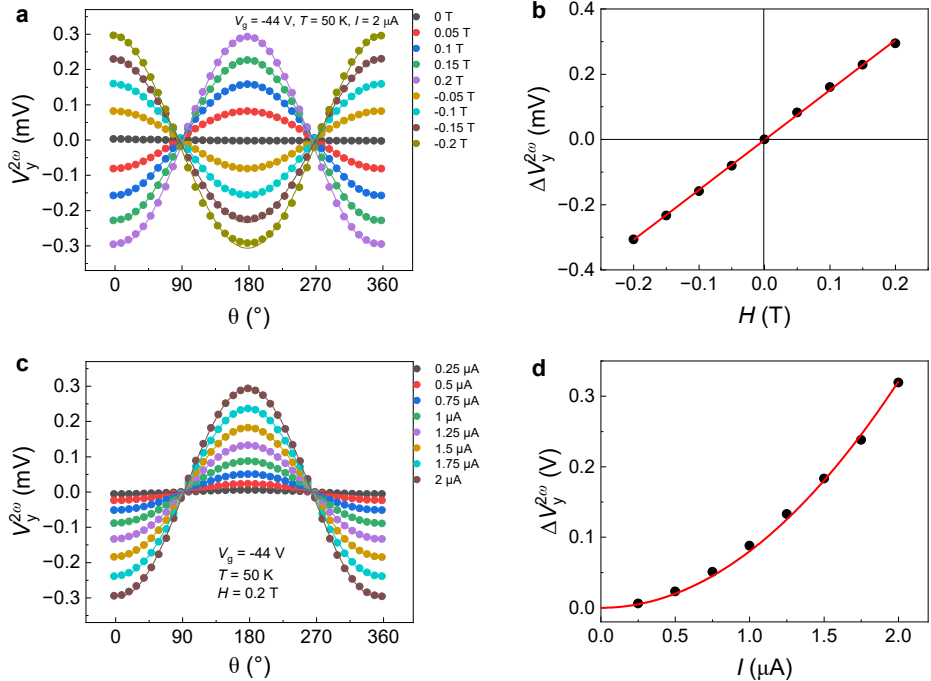

**Supplementary Fig. 7| The magnetic field θ, strength $H$ and current dependences of $V_y^{2\omega}$ in moiré superlattice Device 3.** **a**, $V_y^{2\omega}$ as a function of θ for several different magnetic field strengths at $I$ = 2 μA. A vertical offset has been subtracted from the data to better visualize the field-driven effect. **b**, $\Delta V_y^{2\omega}$ as a function of magnetic field. The solid line represents a linear fit to the

data. **c**, $V_y^{2\omega}$ as a function of θ for several different currents at $H$ = 0.2 T. A vertical offset has been subtracted from the data to emphasize the field-driven effect. **d**, $\Delta V_y^{2\omega}$ as a function of current. The red line shows a quadratic fit to the data. All results in this figure were measured at $T$ = 50 K and $V_g$ = -44 V.

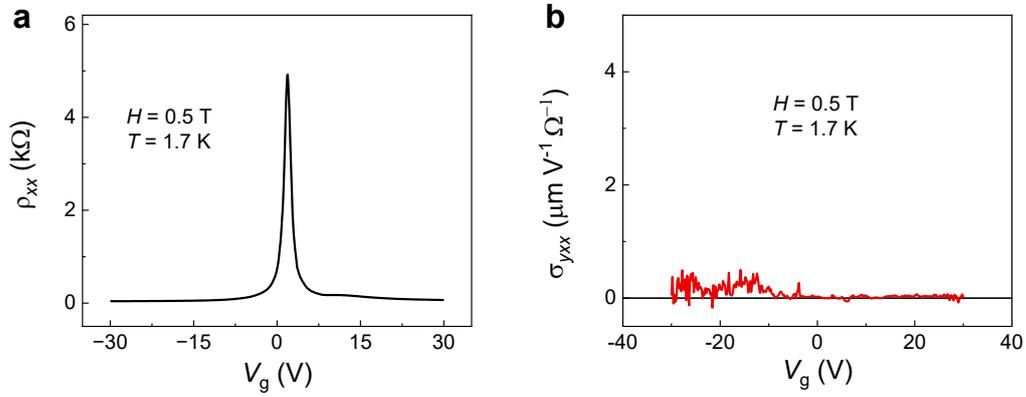

**Supplementary Fig. 8| Negligible field-indued NHE in non-aligned device. a**, The longitudinal resistivity $\rho_{xx}$ as a function of $V_g$, showing no evidence for secondary DP. **b**, The nonlinear Hall conductivity $\sigma_{yxx}$ as a function of $V_g$. These results were obtained at $H$ = 0.5 T and $T$ = 1.7 K.